# Using Google Ngram Viewer for Scientific Referencing and History of Science

A.C. Sparavigna[1] and R. Marazzato[2]

1 Department of Applied Science and Technology, Politecnico di Torino, Torino, Italy
2 Visiting Staff at Department of Mathematical Sciences, Politecnico di Torino, Torino, Italy

**Abstract**: Today, several universal digital libraries exist such as Google Books, Project Gutenberg, Internet Archive libraries, which possess texts from general collections, and many other archives are available, concerning more specific subjects. On the digitalized texts available from these libraries, we can perform several analyses, from those typically used for time-series to those of network theory. For what concerns time-series, an interesting tool provided by Google Books exists, which can help us in bibliographical and reference researches. This tool is the Ngram Viewer, based on yearly count of n-grams. As we will show in this paper, although it seems suitable just for literary works, it can be useful for scientific researches, not only for history of science, but also for acquiring references often unknown to researchers.

## Introduction

Our knowledge capabilities are experiencing a revolution comparable to the one originated by the printing press, introduced in the Western world around 1440 by Johannes Gutenberg, who adapted screw presses and other existing technologies, to create a printing system. The printing devices led to the first mass production of books in Europe [1]. Today, the web search engines, which are part of the Digital Revolution, are transforming manuscripts and printed resources into digital resources, immediately available all over the world. In a certain manner, acting like Gutenberg, the search engines are adapting digital computers, digital records and web providers for disseminating knowledge and culture from books of all literary and scientific disciplines.

Several universal digital libraries exist: among them we find the Google Books, Project Gutenberg, Internet Archive libraries, and many others which concern more specific subjects. Well-known is Google Books, which started as Google Books Search. On the digitalized texts available from these libraries, we can perform several analyses as previously discussed, [2-5], from those pertaining to the time-series analysis to those typically used by network theory and graphs. For what concerns time-series, an interesting tool provided by Google exists, which can help us in bibliographical and reference researches. As we will show in this paper, although it seems suitable just for literary works, it can be useful for scientific researches, not only for the history of science, but also for acquiring references unknown to researchers. This tool is the Ngram Viewer.

## Ngram Viewer

The Ngram Viewer provided by Google is an online viewer, that charts frequencies of any word or short sentence using the yearly count of n-grams found in the sources printed between 1500 and 2015 [6] in American English, British English, French, German, Spanish, Russian, Hebrew, and Chinese. The n-grams are matched by case-sensitive spelling, comparing exact letters and plotted on a graph [6]. The viewer was developed by J. Orwant and W. Brockman and released in mid-December 2010; it was inspired by a prototype created by J.-B. Michel and E. Aiden from Harvard's Cultural Observatory and Yuan Shen from MIT and S. Pinker [6].

In the fields of computational linguistics, an n-gram is a contiguous sequence of n-items from a given sequence of text or speech. The items can be phonemes, syllables, letters, words or base pairs according to the application. An n-gram of size 1 is referred to as a "unigram". Then we have the size 2 "bigram", size 3 "trigram" and so on. The n-gram models are widely used in statistical natural language processing [7,8]. In speech recognition, phonemes and sequences of phonemes are modeled using a n-gram distribution [9].

As discussed in [9], the idea can be traced to an experiment by Claude Shannon's work in information theory. Shannon posed the question: given a sequence of letters (for example, the sequence "for ex"), what is the likelihood of the next letter? From training data, one can derive a probability distribution for the next letter given a history of size n: a = 0.4, b = 0.00001, c = 0, and so on. The probabilities of all possible "next-letters" sum to 1 of course. In [6], it is reported that research based on the n-gram database has included the finding of "correlations between the emotional output and significant events in the 20th century, such as World War II", [10], or to check and challenge popular trend statements of modern societies [11]. The item of Wikipedia concludes that Google Viewer represents a valuable research tool for digital humanities. Let us stress that the graphs proposed by the Viewer are time-series. Detailed discussion of the tool is given by Google at https://books.google.com/ngrams/info.

In the following Figure 1, we can see a screenshot of a graph concerning names proposed as an example by the Viewer (Albert Einstein, Sherlock Holmes and Frankenstein).

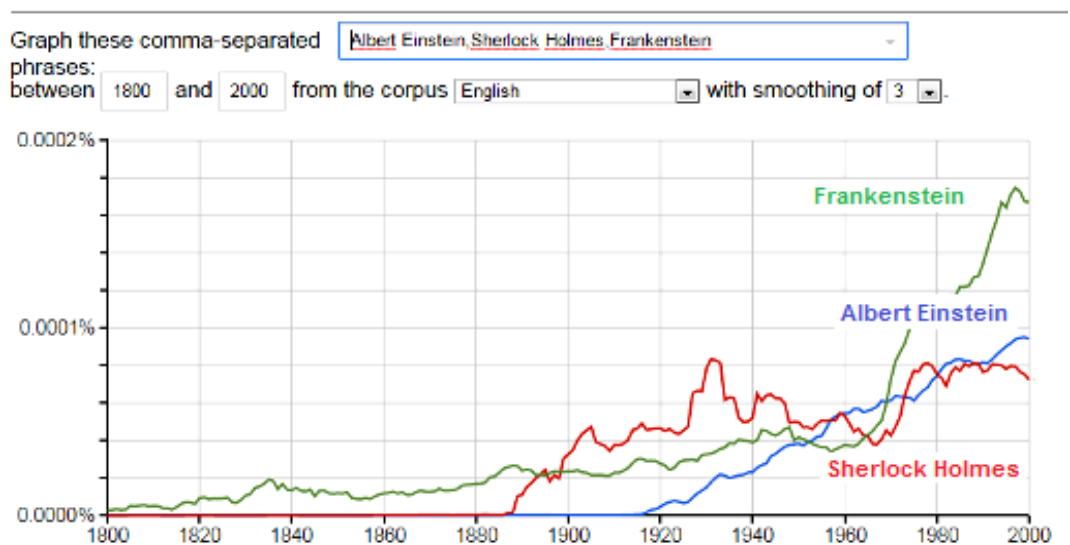

*Figure 1: Screenshot of the tool from Google Ngram site, https://books.google.com/ngrams*

For what concerns our following analysis, let us consider two points for the Google web page, discussing the Viewer. If we start our research from old documents, it happens that graphs are giving more spikes and plateaus in early years. The answer is this: publishing was a relatively rare event in the 16th and 17th centuries. If a phrase occurs in one book in one year but not in the preceding or following years, that creates a taller spike than it would in later years. The other question is: "Many more books are published in modern years. Doesn't this skew the results?" The answer tells us that the results are normalized. "It would if we didn't normalize by the number of books published in each year".

Let us start discussing some examples from literary works and then we will show how the Viewer can be used for scientific researches too.

**A literary case: Frankenstein character**

Let us start from one of the names we find in Figure 1. Frankenstein is a novel written by Mary Wollstonecraft Shelley, about a young science student Victor Frankenstein, who creates a creature in an unorthodox scientific experiment. The first edition of the novel was published anonymously in London in 1818. Shelley's name appears on the second edition, published in France in 1823.

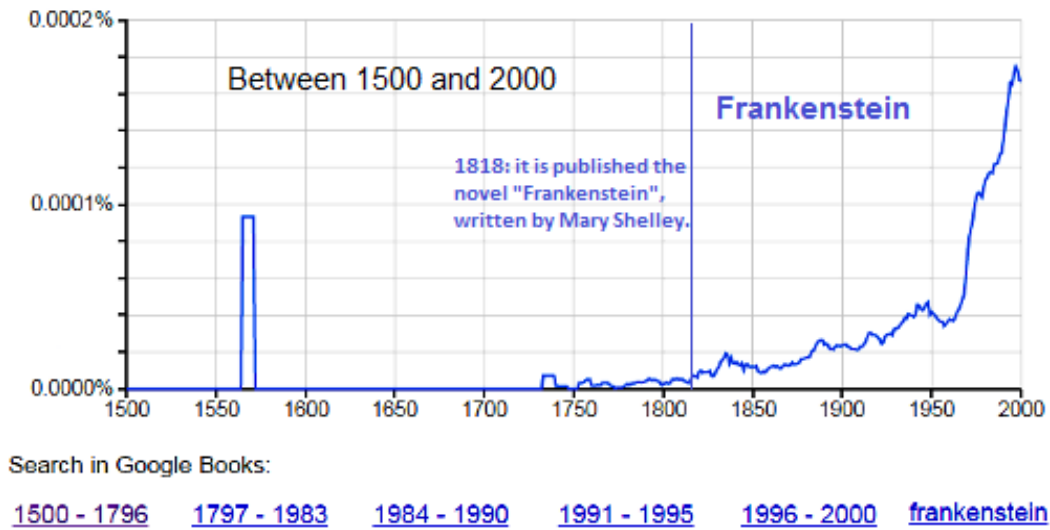

*Figure 2: The time-series corresponding to Frankenstein.*

Mary Shelley explained that she derived the name Frankenstein from a dream-vision. Despite her statement, some sources have been suggested some possible inspiration for the name of this character. The German name Frankenstein means "stone of the Franks", and it is associated with various places in Germany, including the Frankenstein Castle in Darmstadt and another castle in Frankenstein, a town in the Palatinate. In [12], it is argued that Mary Shelley could have visited the Frankenstein Castle near Darmstadt in 1814 during a travel from Switzerland to England. It was at this castle that Conrad Dippel was born, an alchemist that experimented with human bodies. In fact, if we observe the Figure 2, which considers a time-series from 1500, we see that we have a small spike around 1730 and a large one around 1570. Below the graph, a form is displayed for searching the books in several time intervals. And then, using that corresponding to 1500-1796, we find several books; the first one returned is shown by the screenshot of Figure 3. The book tells that Marshal Neuperg had marched toward Frankenstein, which is the town of the above-mentioned castle. The book was published in 1789. The spike about 1730 concerns the town of Frankenstein too.

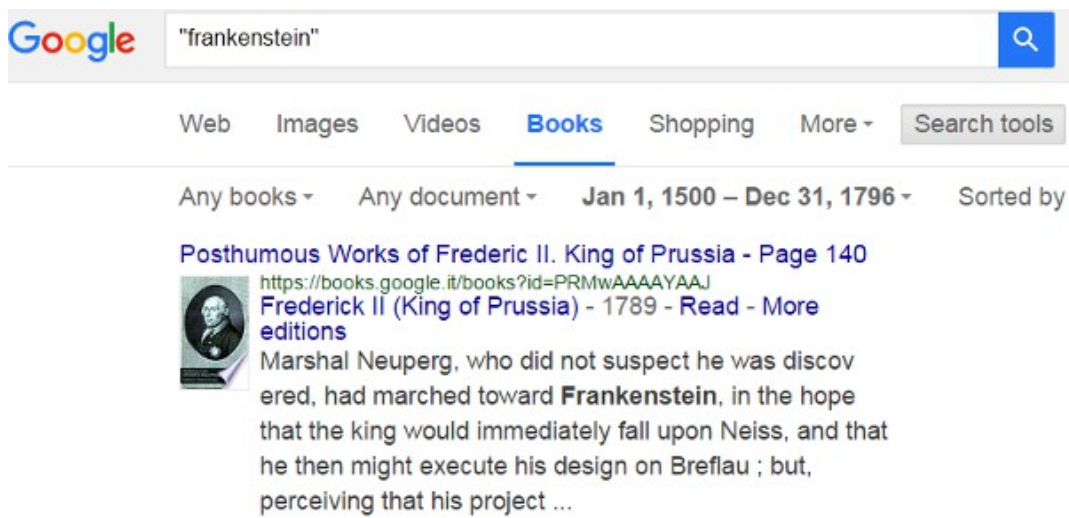

*Figure 3: The book mentioning the town of Frankenstein.*

However, in the Figure 2, we see a large spike in the time-series about 1570. This is due to the fact that one of the books in the collection of the related time interval shows a wrong Roman number representing its print year. Then, this is not a problem of the Viewer, but a problem of the printed book. It is: "Essays on the Christian Aspect of Some Pressing Problems", Reid, 1594, because, inside the text we find MDXCIV, instead of 1894. By the way, in the text the character of Shelley's novel is mentioned: "... think that if it were to be transfigured before us into hard concrete facts, we would shrink back in dismay from the Frankenstein we had raised."

For what concerns a general analysis of the role of Frankenstein in horror and Gothic literature, we could try to find the most favored character. Here, in the following plot (Figure 4), obtained using Google Ngram Viewer, we have the answer. The Vampire wins, and in the plot we can also see the effect of the series of Twilight novels.

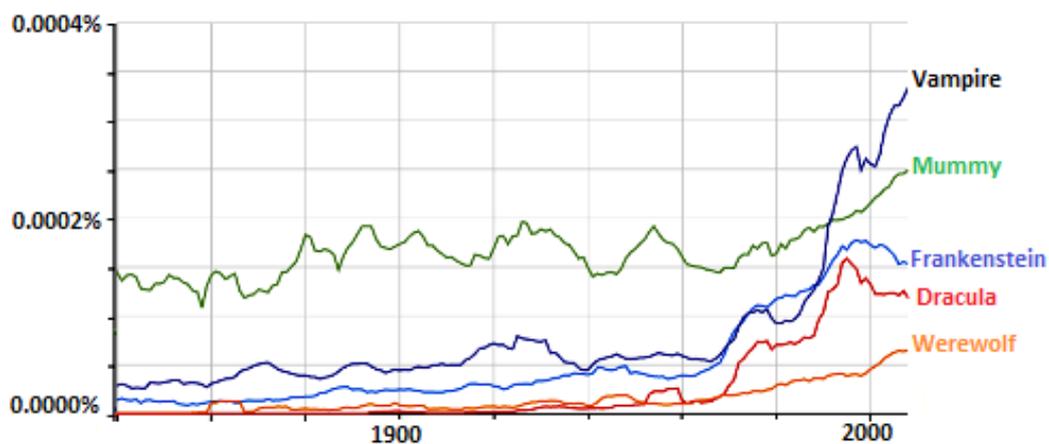

*Figure 4: Google Ngram Viewer tells us the most favored character, among those we are considering. The Vampire wins, and in the plot we can see also the effect of Twilight novels.*

Another possibility is to consider some literature for children. Here the result in the following Figure 5. The behavior of Harry Potter character is quite interesting.

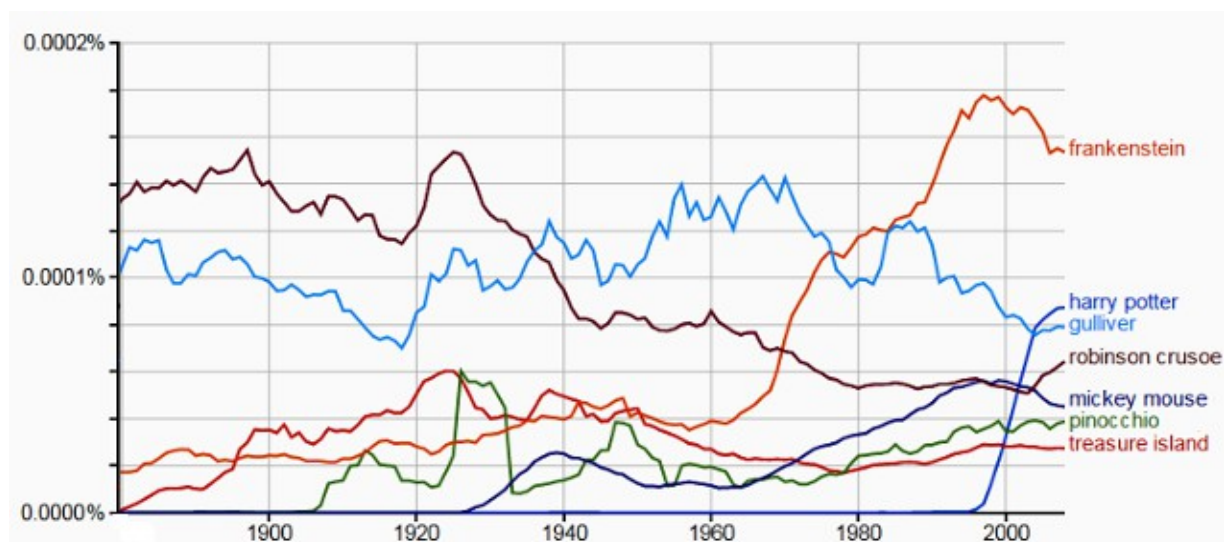

*Figure 5: In this time-series, Google Ngram Viewer is used to compare some literature for children. Note the interesting behavior of Harry Potter.*

**Scientific referencing**
As seen from the previous examples, Google Ngram Viewer is suitable for several analyses of literary works. However, it is quite interesting for scientific researches too, and not only for the history of science, but also for acquiring references unknown to researchers. Of course, we can compare the impact on literature of several scientists, as in the Figure 1, which reports the data for Albert Einstein's name. In this manner, it is easy to find their popularity. However, in this paper, our aim is not only the use of the Viewer in the field of history of science and of its sociology, but also the capability of it of finding new references concerning some specific fields of science. In the following examples we will propose some cases from condensed matter physics.

**Liquid crystals**
In the following Figure 6, we see a quite interesting time-series of adjectives concerning the mesophases of liquid crystals. In the physics of condensed matter, a mesophase is an intermediate state of matter between liquid and solid. Georges Friedel, in 1922, called attention to these mesomorphic states in his paper on liquid crystals [13]. These are certain organic materials that "do not show a single transition from solid to liquid, but rather a cascade of transitions involving new phases". The mechanical properties and the symmetry properties of these phases are intermediate between those of a liquid and those of a crystal. For this reason they have often been called liquid crystals. A more proper name is 'mesomorphic phases' [14].

Let us consider, in the Figure 6, the plot concerning 'smectic'. This adjective is related to a phase of a liquid crystal characterized by arrangement of molecules in layers, with the long axes of all the molecules in a given layer being parallel. The axes can be perpendicular or slightly inclined to the plane of the layer. In http://www.merriam-webster.com/dictionary/smectic, we find that this term comes from the Latin *smecticus*, which means 'cleansing, having the properties of soap', from Greek *smēktikos*, from *smēchein* to clean. It is also told that the year of the first known use is 1923.

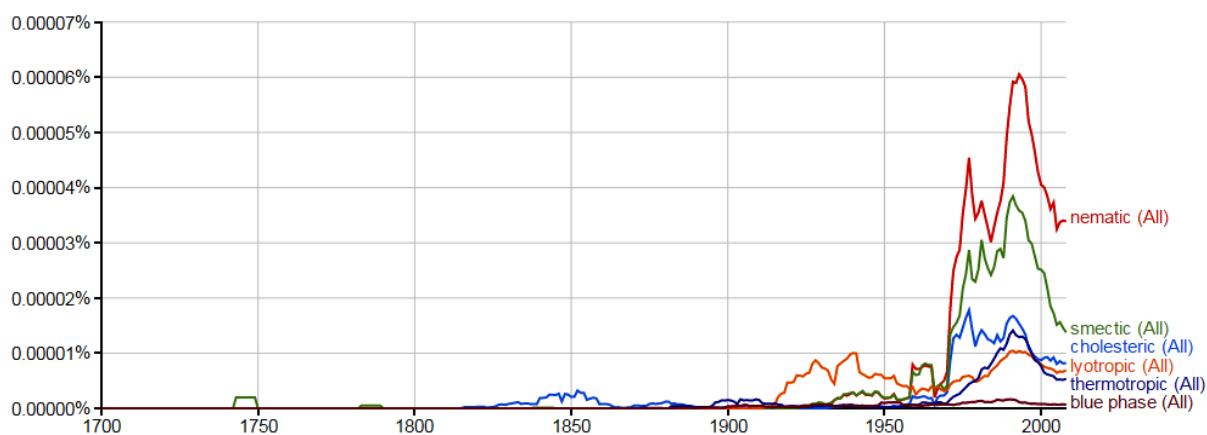

*Figure 6: Some adjectives linked to the mesophases of liquid crystals.*

Now, in the graph, we see a spike for 'smectic' about 1750. This is not a mistake of the Viewer. It is a recipe of an 'Emplastrum Smecticum', made of Red-Lead, oil of olives, Venice soap and yellow wax, in the *Pharmacopœia officinalis & extemporanea*: or, a Complete English dispensatory, by John Quincy, 1730, Osborn and Longman in London. Let us consider the adjective 'cholesteric' too: in http://www.merriam-webster.com/dictionary/cholesteric, it is told that it is related to the "phase of a liquid crystal characterized by arrangement of molecules in layers with the long molecular axes parallel to one another in the plane of each layer and incrementally displaced in successive layers to give helical stacking". From the same site, we find that 'cholesteric' is related to 'cholesterol', and that the first known use was in 1942. However, the plot tells us of its previous use, and then, from

Google Books, we have that, in a "Dictionary of Chemistry", by Andrew Ure, published by Robert Desilver, Philadelphia, 1821, we find the Cholesteric Acid.

Besides these curiosities, we can see from the plots the general trend of the researches concerning liquid crystals, which is decreasing during the last years.

**Carborundum and Silicon Carbide**

Silicon carbide (SiC) and carborundum are the same compound of silicon and carbon. The powder of this compound has been mass-produced since 1893 to be used as an abrasive. It was known with the trade name Carborundum [15]. Grains of silicon carbide can be sintered to form very hard ceramics, widely used in applications requiring high endurance. Electronic applications of SiC for light-emitting diodes (LEDs) and detectors were first demonstrated around 1907 [16]. SiC is used in semiconductor devices that operate at high temperatures and that, for this reason, require a high thermal conductivity, a property possessed by this material [17].

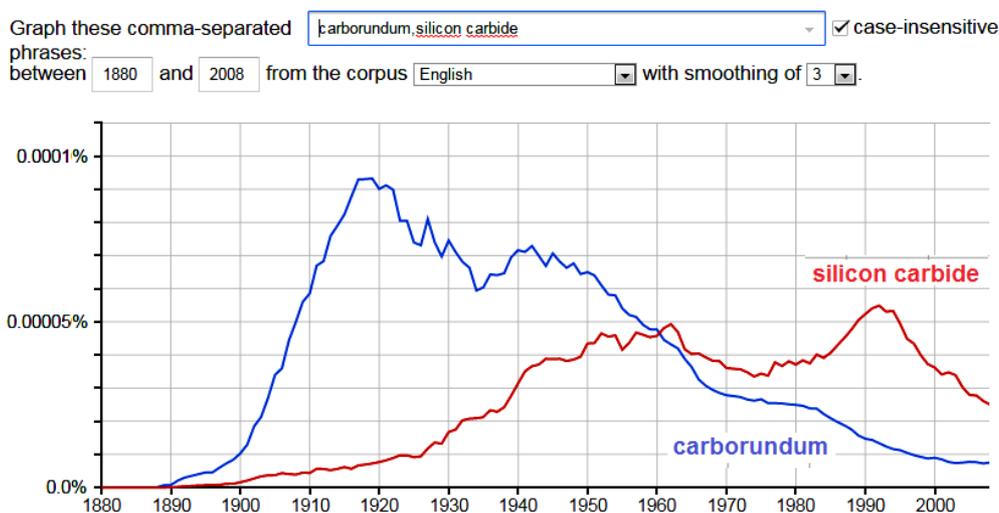

*Figure 7: The same compound has two names depending on the use made of it.*

Let us consider the two terms used for the same compound in the Ngram Viewer, the result is given in the Figure 7. We can see that the two names, corresponding to a different use of the same compound have a different graph. Among the first books on Carborundum, that we can find using the form of the viewer, we have the book entitled "Carborundum: Its History, Manufacture and Uses", by Edward Goodrich Acheson, 1893. Acheson (1856-1931) was an American chemist, inventor of the Acheson process, which is still used to produce carborundum [18], and later a manufacturer of carborundum and graphite. In [19], it is described the process: "when a mixture of silica and carbon is heated to a sufficiently high temperature the silica is reduced, and if a proper amount of carbon is present silicon carbide is formed. At relatively low temperatures the silica-carbon mixture yields a greenish-colored amorphous substance, but as the temperature is increased this is converted into crystalline silicon carbide, which has been named 'carborundum' by its discoverer, Mr. E.G. Acheson. If the temperature is increased considerably beyond that of its formation, carborundum is decomposed, the silicon being expelled as vapor and the carbon left behind in the form of graphite" [19]. Acheson was granted a patent for graphite manufacture in 1896, and commercial production started in 1897.

As a consequence of our research on Ngram Viewer, we learned a link between silicon carbide and graphite. Since we talked about silicon, let us consider also a time-series concerning silicon and some other materials (diamond, aluminum, germanium and gallium). It is shown in the Figure 8.

Note that the role of silicon in books has today a decreasing trend. If we consider diamond as a reference material, we have a period of twenty years during which silicon was more important, this was the age of silicon.

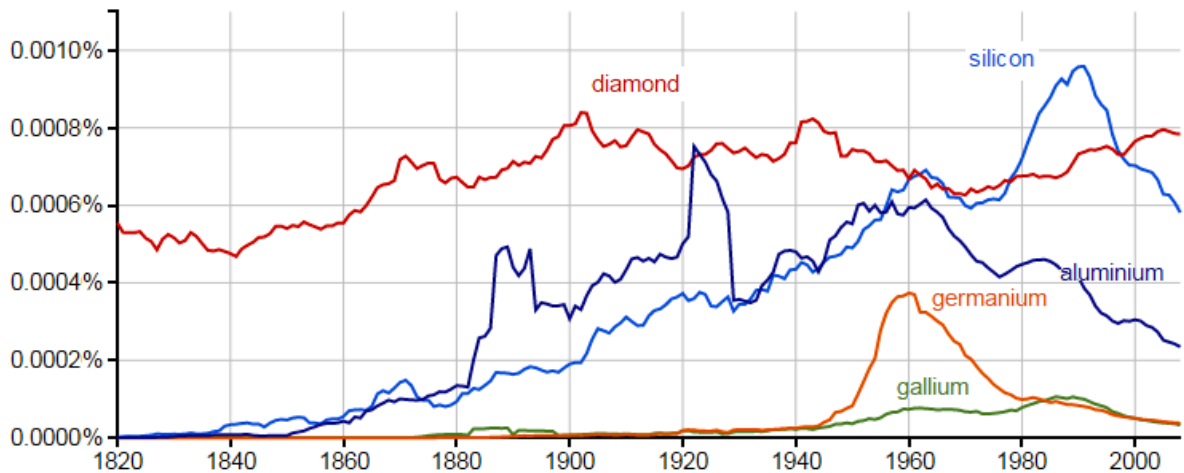

*Figure 8: Some materials and the age of Silicon, when it was more important than diamond.*

Let us conclude with a time-series concerning nanotubes, fullerene, carbon fibres and graphene, shown in the Figure 9. It could be surprising, but fullerene is more popular than graphene. In fact, graphene was considered in literature since 1985.

Graphene is a two-dimensional honeycomb arrangement of carbon atoms. This material is considered so significant that it earned Russian scientists Andre Geim and Konstantin Novoselov 2010 Nobel Prize in Physics. As explained in [20], "in 2003, one ingenious physicist took a block of graphite, some Scotch tape and a lot of patience and persistence and produced a magnificent new wonder material that is a million times thinner than paper, stronger than diamond, more conductive than copper. It is called graphene, and it took the physics community by storm when the first paper appeared the following year". So the discovery of graphene is related to that of the free-standing graphene.

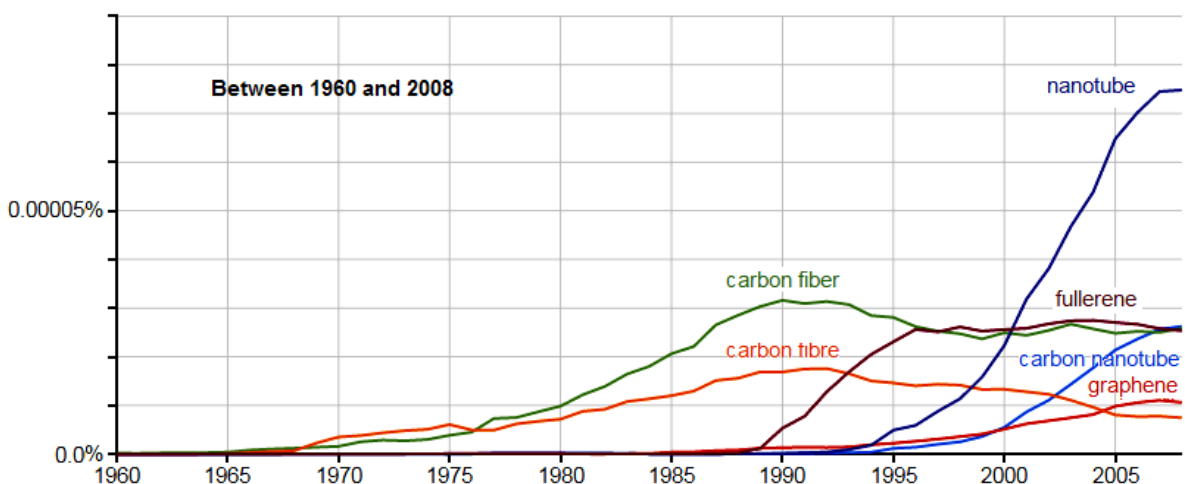

*Figure 9: In 2008, fullerene was most popular than graphene.*

The fact that graphene is so important for new physics is not evident from the time-series of Figure 9. This could happen because it is not yet so popular as the carbon nanotubes. An interesting evolution of the Google Books Ngram Viewer could be an analog tool for Google Scholar. This would be more focused on research, and could probably provide more confident results for scientific subjects.